\documentclass[aps,prl,reprint,superscriptaddress,twocolumn,showkeys,amsmath,amssymb,longbibliography,floatfix]{revtex4-2}

\usepackage[colorlinks,linkcolor=blue,citecolor=blue,urlcolor=blue]{hyperref}

\usepackage{textgreek}

\usepackage{multirow}
\usepackage{subfigure}
\usepackage{color}
\usepackage{mathrsfs}
\usepackage{hyperref}
\usepackage[normalem]{ulem}
\usepackage{bm}

\usepackage{amssymb}   
\usepackage{amsmath}
\renewcommand\vec[1]{\ensuremath\boldsymbol{#1}} 

\usepackage{amsfonts, relsize, color}
\usepackage{graphics}
\usepackage{graphicx}
\usepackage{subfigure}
\usepackage{hyperref}
\usepackage{color}
\usepackage{comment}

\begin{document}

\title{Topologically distinct atomic insulators}

\author{Sanjib Kumar Das}
\affiliation{Department of Physics, Lehigh University, Bethlehem, Pennsylvania, 18015, USA}

\author{Sourav Manna}
\affiliation{Department of Condensed Matter Physics, Weizmann Institute of Science, Rehovot 7610001, Israel}
\affiliation{Raymond and Beverly Sackler School of Physics and Astronomy, Tel-Aviv University, Tel Aviv 6997801, Israel}

\author{Bitan Roy}
\affiliation{Department of Physics, Lehigh University, Bethlehem, Pennsylvania, 18015, USA}

\begin{abstract}
Topological classification of quantum solids often (if not always) groups all trivial atomic or normal insulators (NIs) into the same featureless family. As we argue here, this is not necessarily the case always. In particular, when the global phase diagram of electronic crystals harbors topological insulators with the band inversion at various time-reversal invariant momenta ${\bf K}^{\rm TI}_{\rm inv}$ in the Brillouin zone, their proximal NIs display noninverted band-gap minima at ${\bf K}^{\rm NI}_{\rm min}={\bf K}^{\rm TI}_{\rm inv}$. In such systems, once topological superconductors nucleate from NIs, the inversion of the Bogoliubov de Gennes bands takes place at ${\bf K}^{\rm BdG}_{\rm inv}={\bf K}^{\rm NI}_{\rm min}$, inheriting from the parent state. We showcase this (possibly general) proposal for two-dimensional time-reversal symmetry-breaking insulators. Then distinct quantized thermal Hall conductivity and responses to dislocation lattice defects inside the paired states (tied with ${\bf K}^{\rm BdG}_{\rm inv}$ or ${\bf K}^{\rm NI}_{\rm min}$), in turn unambiguously identify different parent atomic NIs.          
\end{abstract}

\maketitle

\textbf{\textit{Introduction}}.~The world of insulators fragments into two sectors according to the topology and geometry of the bulk electronic wavefunction in quantum crystals: topological insulators (TIs) and normal insulators (NIs)~\cite{Hasan2010, Qi2011}. TIs manifest bulk-boundary correspondence, featuring robust gapless modes at crystal interfaces, such as edge, surface, corner and hinge, for example. When combined with the crystal symmetry, the family of TIs hosts a rich fair showcasing strong, weak, crystalline, higher-order, and atomically obstructed TIs~\cite{kanemele2006, BHZ2006, Fukane2007, moorebalents2007, rahulroy2009, tenfold2010, tenfoldPRB2008, vanderbilt2011, Fu2011, Slager2012, Shiozaki2014, BBH2017, BBHPRB2017, bernevig2017, Vishwanath2017NatComm, ashvin2018, Calugaru2019, Zhang2019, Vergniory2019, Tang2019, khalaf2021}. By contrast, atomic or normal insulators, although abundant in nature, do not accommodate any gapless topological boundary modes. Naturally, within the topological classification scheme of quantum materials, NIs are grouped into a single featureless family. A question, therefore, can be raised. \emph{Can we topologically distinguish such NIs?} leaving aside their nontopological spectroscopic characterization based on the band gap minima momenta (${\bf K}^{\rm NI}_{\rm min}$).

\begin{figure}[t!]
	\includegraphics[width=0.90\linewidth]{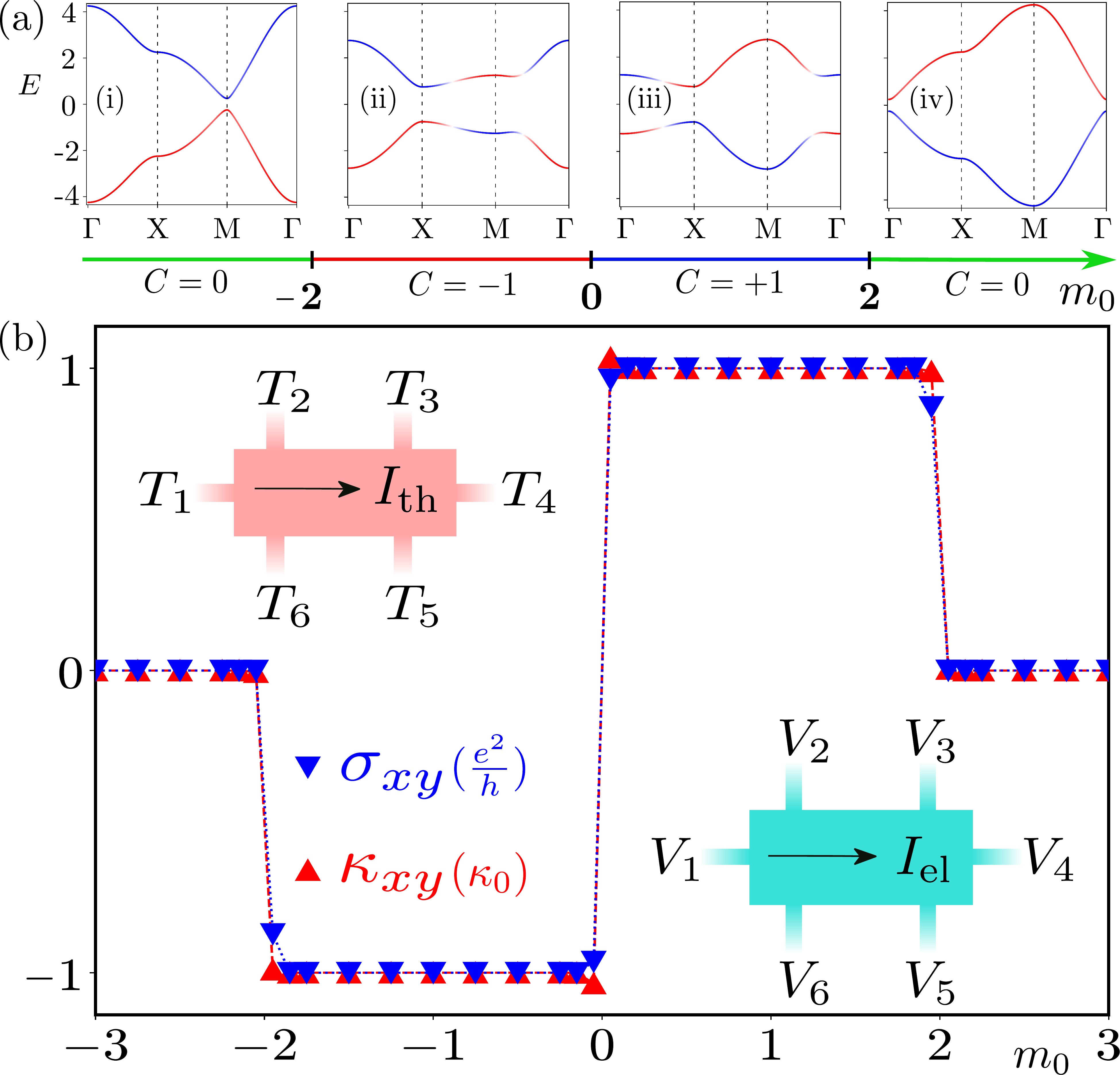}
	\caption{(a) Phase diagram of the normal-state Hamiltonian [Eq.~\eqref{eq:qwz_ham}] in terms of the Chern number $C$ [Eq.~\eqref{eq:chernnumber}] for $t=t_0=1$. In each insulating phase, the band structure displays parity polarization of the eigenvectors in red ($+$) and blue ($-$) for (i) $m_0=-2.25$, (ii) $m_0=-0.75$, (iii) $m_0=0.75$ and (iv) $m_0=2.25$. Bands are noninverted (inverted) in NIs (TIs). Here, we follow the path $\Gamma \to {\rm X} \to {\rm M} \to \Gamma$ in the BZ. (b) The six-terminal electrical ($\sigma_{xy}$) and thermal ($\kappa_{xy}$) Hall conductivities as a function of $m_0$, computed in a rectangular system (see the insets) of length $L=200$ and width $W=100$. In TIs, both $\sigma_{xy}=C$ and $\kappa_{xy}=C$ [in units of $\kappa_0=\pi^2 k^2_B T/(3h)$] at $T=0.01$. Dotted lines are guide to the eye. 
	}~\label{fig:pd}
\end{figure}

\begin{figure}[t!]
	\includegraphics[width=0.90\linewidth]{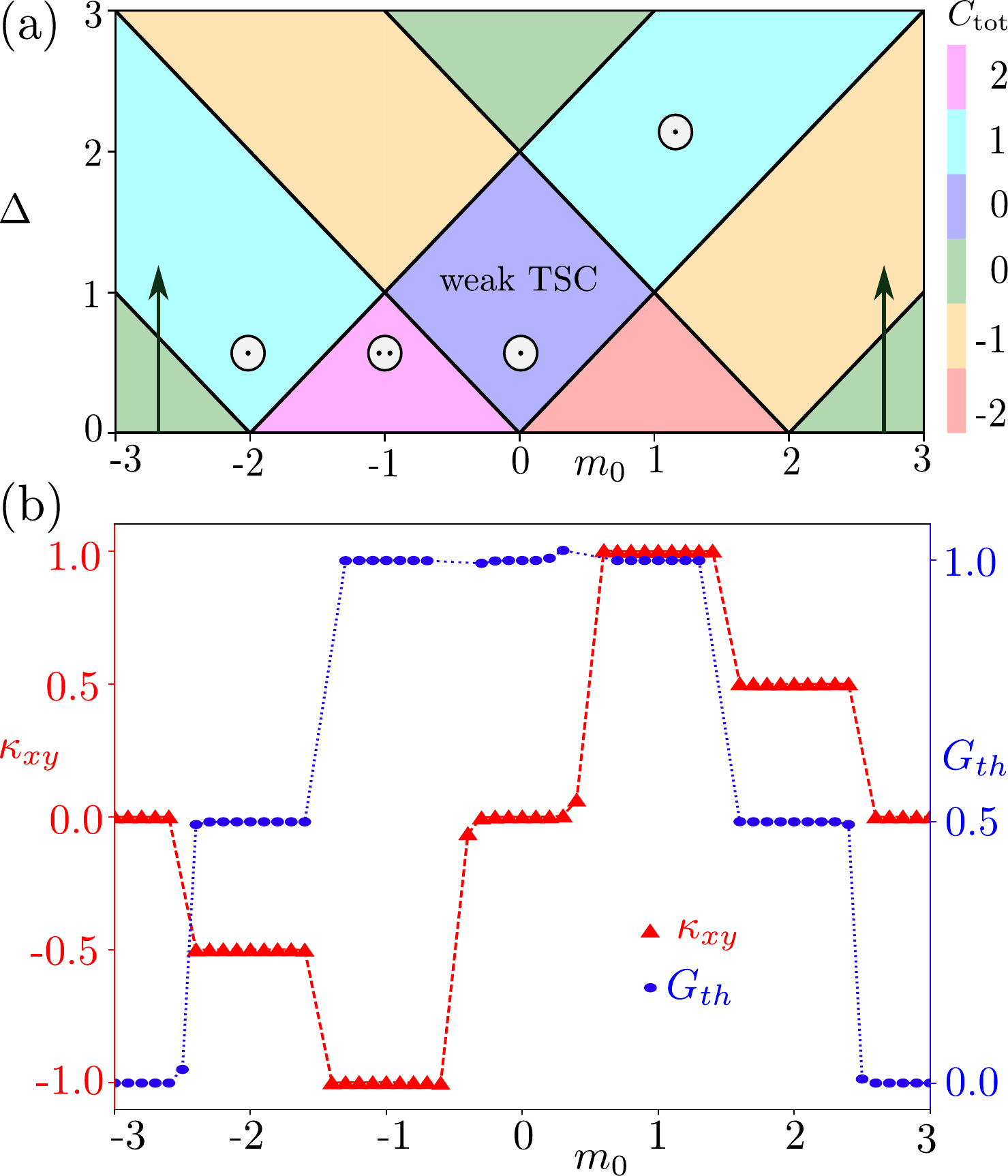}
	\caption{(a) Phase diagram of $\mathcal{H}_{\rm BdG}(\vec{k})$ [Eq.~\eqref{eq:Nambu_ham}]. Phases are colored according to the total Chern number ($C_{\rm tot}$) and `weak TSC' possesses a weak invariant, the Zak phase. Circled phases with single (double) dot(s) support one (two) pair(s) of dislocation modes [Fig.~\ref{fig:ldos}]. (b) Thermal Hall conductivity ($\kappa_{xy}$) and longitudinal thermal conductance ($G_{th}$) as a function of $m_0$ for $\Delta=0.5$, computed in a system of $L=2 W=80$ [Fig.~\ref{fig:pd}(inset)] at $T=0.01$. In units of $\kappa_0$, $\kappa_{xy}=C_{\rm tot}/2$ and $G_{th}=|k_{xy}|$. But, in the weak TSC phase $G_{th}=1$ (in units of $\kappa_0$). Arrows show two TSCs resulting from two NIs, distinguished from the paired state responses [(b) and Fig.\ref{fig:ldos}], confirming ${\bf K}^{\rm BdG}_{\rm inv}={\bf K}^{\rm NI}_{\rm min}$ [Fig.~\ref{fig:bs}].  
	}~\label{fig:pd_thc}
\end{figure}

\begin{figure}[tb]
\includegraphics[width=0.95\linewidth]{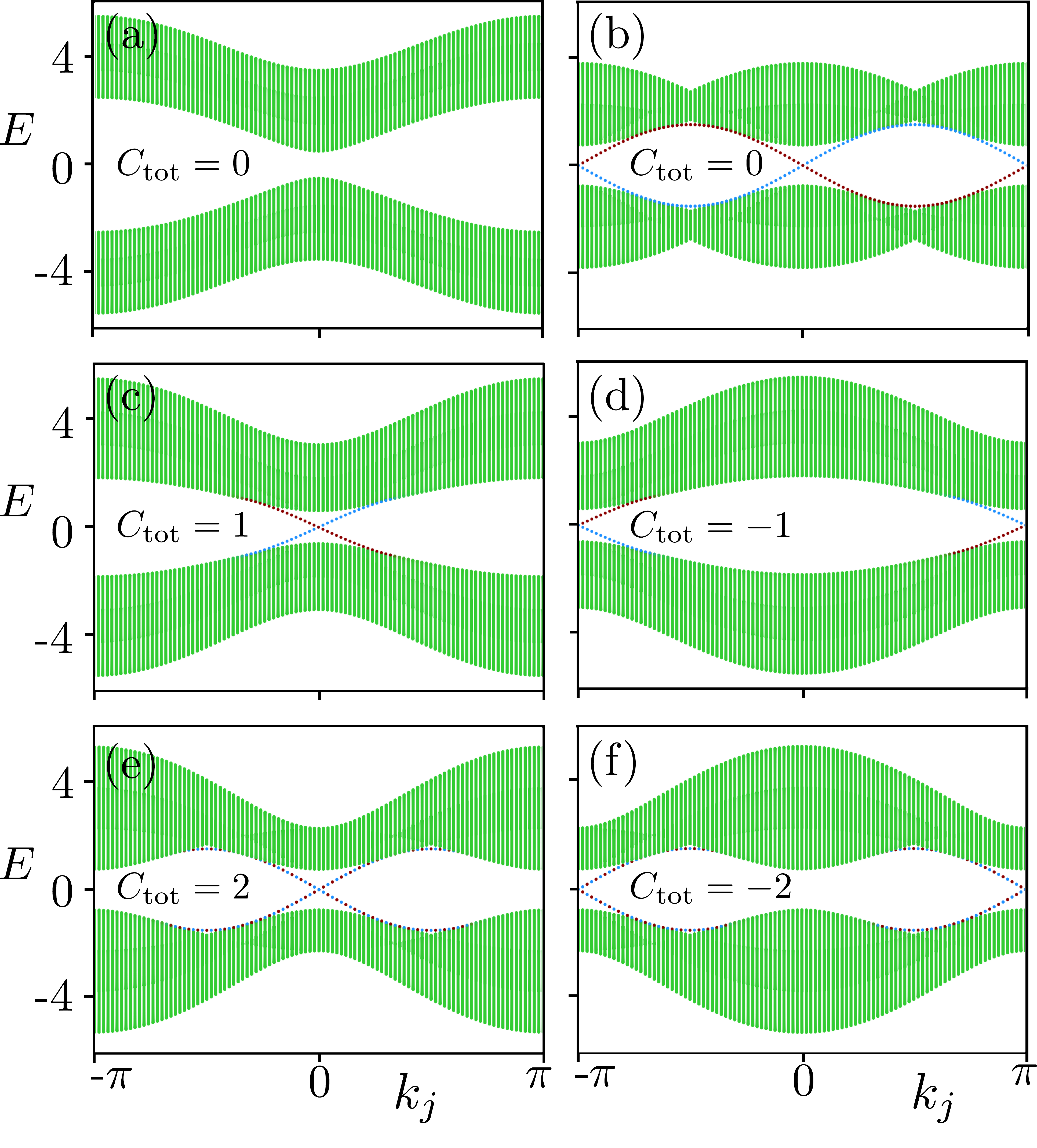}
\caption{Band structure of ${\mathcal H}_{\rm BdG}(\vec{k})$ [Eq.~\eqref{eq:Nambu_ham}] in a semi-infinite system with $k_j$, where $j=x$ or $y$ and $120$ unit cells in the $y$ or $x$ direction for $\Delta=0.5$. The values of $m_0$ are (a) $3.0$, (b) $0.0$, (c) $2.0$, (d) $-2.0$, (e) $1.0$, and (f) $-1.0$. The total Chern number ($C_{\rm tot}$) is quoted in each panel [Fig.~\ref{fig:pd_thc}(a)]. In (b), counter-propagating edge modes result from a weak invariant (Zak phase). To display doubly degenerate edge modes in (e) and (f), we plot one of them for odd and the other one for even momentum grids. Red (blue) and green colors indicate states that are localized on the left (right) edge and in the bulk of the system, respectively.}
\label{fig:bs}
\end{figure}

\begin{figure*}[tb]
\includegraphics[width=0.95\linewidth]{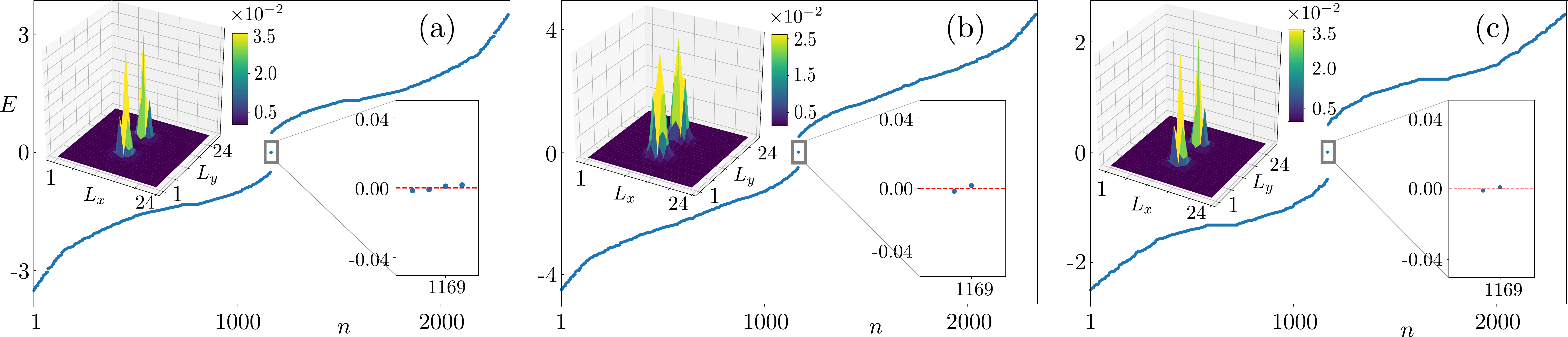}
\caption{Energy spectra of ${\mathcal H}_{\rm BdG}(\vec{k})$ [Eq.~(\ref{eq:Nambu_ham})] in the presence of an edge dislocation-antidislocation pair with Burgers vectors ${\bf b}= \pm a \hat{\bf e}_x$, placed symmetrically in a periodic system with linear dimensions $L=24$ in the $x$ and $y$ directions, for $\Delta=0.5$, and (a) $m_0=-1.0$, (b) $m_0=-2.0$, and (c) $m_0=0.0$, yielding $C_{\rm tot}=+2$, $+1$, and $0$ (with nontrivial Zak phase) [Fig.~\ref{fig:pd_thc}(a)], respectively. The insets show near zero energy states, whose local density of states is highly localized around the defect cores.}
\label{fig:ldos}
\end{figure*}

Here, we provide an \emph{indirect} affirmative answer to this question by considering a paradigmatic toy square lattice model for two-dimensional (2D) time-reversal symmetry- (${\mathcal T}$-) breaking insulators~\cite{Qi2006}. We show if the global phase diagram of quantum materials supports TIs featuring the hallmark band inversion at different time-reversal invariant momenta ${\bf K}^{\rm TI}_{\rm inv}$ in the Brillouin zone (BZ), then, their respective proximal NIs display a band-gap minima at ${\bf K}^{\rm NI}_{\rm min}$=${\bf K}^{\rm TI}_{\rm inv}$ [Fig.~\ref{fig:pd}]. In such systems, when topological superconductors (TSCs) nucleate from NIs [Fig.~\ref{fig:pd_thc}(a)], the inversion of the Bogoliubov de Gennes (BdG) bands takes place at ${\bf K}^{\rm BdG}_{\rm inv}={\bf K}^{\rm NI}_{\rm min}$ [Fig.~\ref{fig:bs}]. Although half-quantized thermal Hall conductivity ($\kappa_{xy}$) reveals the topological nature of the paired states [Fig.~\ref{fig:pd_thc}(b)], dislocation lattice defects, sensitive to ${\bf K}^{\rm BdG}_{\rm inv}$, in turn, underpins ${\bf K}^{\rm NI}_{\rm min}$ [Fig.~\ref{fig:ldos}]. Therefore, responses of TSCs allow us to identify and distinguish their parent NIs. Specifically, when a TSC, characterized by a half-quantized $\kappa_{xy}$, stems from a NI with the band-gap minima at a finite momentum, only then robust zero-energy localized Majorana modes appear near the dislocation core. We present a simple mathematical proof to generalize this proposal to arbitrary dimensions (larger than one) and symmetry class to classify NIs from the responses of their proximal TSCs, operative under the only assumption that a half-filled system always describes an insulator.

\textbf{\textit{Normal state}}.~The Hamiltonian for 2D ${\mathcal T}$-breaking insulators on a square lattice reads $H=\sum_{\vec{k}} \Psi^\dagger_{\vec{k}} {\mathcal H}(\vec{k}) \Psi_{\vec{k}}$, where $\Psi^\top_{\vec{k}}=[c^+_{\vec{k}}, c^-_{\vec{k}}]$, and $c^\tau_{\vec{k}}$ is the fermionic annihilation operator with momentum $\vec{k}$ and parity $\tau=\pm$~\cite{Qi2006}. The $\vec{k}$-dependent operator is given by ${\mathcal H}(\vec{k})={\boldsymbol \tau} \cdot \vec{d}(\vec{k})$ with
\begin{equation}~\label{eq:qwz_ham}
\hspace{-0.20cm}
\vec{d}(\vec{k}) = \bigg( t \sin(k_{x}a), t \sin(k_{y}a), m_0-t_0 \sum_{j=x,y} \cos(k_{j}a) \bigg).
\end{equation}
Vector Pauli matrix ${\boldsymbol \tau}=(\tau_x, \tau_y, \tau_z)$ operates on the parity indices ($\pm$). Throughout, we set $t=t_0=1$, and the lattice constant $a=1$. Then, this model hosts TIs in the regime $-2<m_0<2$, and NIs otherwise. Each TI supports one chiral edge mode, encoding the first Chern number $C=\pm 1$, defined within the first BZ as~\cite{TKNN1982}
\begin{equation}~\label{eq:chernnumber}
C=\int_{\rm BZ} \dfrac{d^{2}{\vec k}}{4\pi} \:\: \big[ \partial_{k_x} \hat{\vec{d}}(\vec{k}) \times \partial_{k_y} \hat{\vec{d}}(\vec{k}) \big] \cdot \hat{\vec{d}}(\vec{k}),
\end{equation}
manifesting the bulk-boundary correspondence, where $\hat{\vec{d}}(\vec{k})=\vec{d}(\vec{k})/|\vec{d}(\vec{k})|$. In NIs, $C=0$. The nontrivial Chern number gives rise to quantized electrical and thermal Hall conductivities, which we discuss shortly. This model breaks the sublattice symmetry ($S$) as there exists no unitary operator that anticommutes with ${\mathcal H}(\vec{k})$ and the ${\mathcal T}$ symmetry~\cite{tenfold2010, tenfoldPRB2008}. A charge-conjugation symmetry (${\mathcal C}$), generated by $\tau_1 {\mathcal K}$ where ${\mathcal K}$ is the complex conjugation~\cite{RoyantiunitaryPhysRevResearch2019}, arises solely because we neglect the particle-hole asymmetry for simplicity as it does not play any role in determining the topology of the insulators.

Various phases of this model Hamiltonian in terms of the Chern number and the associated band structures are shown in Fig.~\ref{fig:pd}(a). The topological regime fragments into two sectors depending on the band inversion momentum in the BZ (${\bf K}^{\rm TI}_{\rm inv}$). Specifically, ${\bf K}^{\rm TI}_{\rm inv}=(0,0)$ ($\Gamma$ point) for $0<m_0<2$, and ${\bf K}^{\rm TI}_{\rm inv}=(\pi,\pi)$ (${\rm M}$ point) for $-2<m_0<0$. In these two phases, $C=+1$ and $-1$, respectively. The transition between them takes place through a band gap closing at the ${\rm X}=(\pi,0)$ or ${\rm Y}=(0,\pi)$ point when $m_0=0$. Two NIs are born from these TIs via bulk gap closings at the ${\rm M}$ and $\Gamma$ points when $m_0=-2$ and $+2$, respectively. Even though the bands are noninverted in NIs, the parity-polarized conduction (valence) band displays band minima (maxima) near the $\Gamma$ and ${\rm M}$ points, respectively, for $m_0>2$ and $m_0<-2$. In this respect, the band-gap minima in NIs occurs at ${\bf K}^{\rm NI}_{\rm min} = {\bf K}^{\rm TI}_{\rm inv}$ of their proximal parent TIs. Although plays no role in topological classification, ${\mathcal H}(\vec{k})$ enjoys an \emph{emergent} inversion symmetry $\tau_z {\mathcal H}(\vec{k}) \tau_z={\mathcal H}(-\vec{k})$, resulting from the opposite parities of two involved orbitals that also pins ${\bf K}^{\rm TI}_{\rm inv}$ and thus ${\bf K}^{\rm NI}_{\rm min}$ at the high symmetry points of the BZ~\cite{SM}, typically the case in topological materials and models~\cite{tenfold2010, tenfoldPRB2008}. Throughout, we assume that there is no translational symmetry breaking causing doubling of unit cell or folding of the BZ. Before addressing the proposal to distinguish NIs with different ${\bf K}^{\rm NI}_{\rm min}$, we characterize the normal state in terms of the electrical ($\sigma_{xy}$) and thermal ($\kappa_{xy}$) Hall responses to facilitate the forthcoming discussion.

\textbf{\textit{Electrical Hall conductivity}}.~We compute $\sigma_{xy}$ in a six-terminal geometry at zero temperature~\cite{SM}. Since mesoscopic details of the device or scattering region and leads play a pivotal role in obtaining meaningful transport responses, here, we briefly discuss their geometry used for the calculations [Fig.~\ref{fig:pd}(b)]. A rectangular scattering region containing the system is maintained at a voltage $V$. It is connected to six terminals. All of them are kept at different voltages with the help of reservoirs. To generate transverse electrical response, we apply a voltage gradient between lead 1 ($V_1=-\Delta V/2$) and lead 4 ($V_4=\Delta V/2$), resulting in a longitudinal electrical current ($I_{\rm el}$) between them. No current is flowing between the transverse leads. They serve as the voltage probes. This setup allows us to calculate $\sigma_{xy}$, generated between the transverse leads by extracting the scattering matrix using Kwant~\cite{Groth2014}. The current-voltage relation is given by ${\bf I}_{\rm el}= {\bf G}_{\rm el} {\bf V}$, with ${\bf I}^\top_{\rm el}=(I_{\rm el},0,0,-I_{\rm el},0,0)$ and ${\bf V}^\top=(-\Delta V/2,V_2,V_3,\Delta V/2,V_5,V_6)$. The conductance matrix ${\bf G}_{\rm el}$ contains only the transmission blocks of the scattering matrix. Upon finding ${\bf G}_{\rm el}$, we extract different voltages from the current-voltage relation. Subsequently, we compute the transverse electrical resistance $R^{\rm el}_{xy}=(V_2+V_3-V_5-V_6)/(2 I_{\rm el})$~\cite{Buttiker1986I, Buttiker1986II, Datta1995}. In units of $e^2/h$, we find $\sigma_{xy}=1/R^{\rm el}_{xy}=C$ [Fig.~\ref{fig:pd}(b)].

\textbf{\textit{Thermal Hall conductivity}}.~The same six-terminal geometry can be used to compute $\kappa_{xy}$. The scattering region is now maintained at a temperature $T$. All six terminals are kept at different temperatures. We apply a temperature gradient between lead 1 ($T_1=-\Delta T/2$) and lead 4 ($T_4=\Delta T/2$). It results in a longitudinal thermal current ($I_{\rm th}$) from lead 1 to lead 4. The current-temperature relation is captured by the matrix equation ${\bf I}_{\rm th}= {\bf A}_{\rm th} {\bf T}$, where ${\bf I}^\top_{\rm th}=(I_{\rm th},0,0,-I_{\rm th},0,0)$ and ${\bf T}^\top=(-\Delta T/2,T_2,T_3,\Delta T/2,T_5,T_6)$. The matrix elements of ${\bf A}_{\rm th}$ are given by~\cite{Long2011, Fulga2020} 
\begin{equation}~\label{eq:currtemp}
A_{{\rm th},ij} =  \int_0^\infty \frac{E^2}{T} \left( -\frac{\partial f(E, T)}{\partial E} \right)\left[\delta_{ij} \mu_j-  {\rm Tr}({\bf t}_{ij}^\dagger {\bf t}_{ij}) \right] dE,
\end{equation}
where $\mu_j$ denotes the number of propagating modes in the $j$th lead, $f(E,T)=1/(1+\exp{[E/(k_{B}T)]})$ is the Fermi-Dirac distribution function, ${\bf t}_{ij}$ is the transmission part of the scattering matrix between the leads $i$ and $j$, and the trace (Tr) is taken over the conducting channels. Upon obtaining ${\bf A}_{\rm th}$, we calculate the temperature at various leads from the current-temperature relation. The transverse thermal resistance is $R^{\rm th}_{xy}=(T_2+T_3-T_5-T_6)/(2I_{\rm th})$. For both electrical and thermal Hall resistances, the average over different terminals is taken to avoid contact resistance effects, giving rise to robust quantized values. Inverting $R^{\rm th}_{xy}$, we obtain $\kappa_{xy}= \left( R^{\rm th}_{xy} \right)^{-1}$~\cite{Read2000, Long2011, Fulga2020, Rego1998}. Notice that the integrand in Eq.~\eqref{eq:currtemp} depends on the derivative of the Fermi-Dirac function, which is valid in the limit $T \rightarrow 0$~\cite{SM}. We compute $\kappa_{xy}$ for $T=0.01$ (in the energy unit). In units of $\kappa_0$, we find $k_{xy}=C$ [Fig.~\ref{fig:pd}(b)].

\textbf{\textit{Superconductivity}}.~Therefore, NIs with distinct ${\bf K}^{\rm NI}_{\rm min}$'s cannot be distinguished from any response of charged fermions. Such a goal can nevertheless be accomplished when the system is conducive to Cooper pairing. The charge-conjugation symmetry allows this system to support only one local pairing~\cite{Sourav2022}. The effective single-particle BdG Hamiltonian then reads $H_{\rm BdG}=\frac{1}{2} \sum_{\vec{k}} \left( \Psi^{\rm Nam}_{\vec{k}} \right)^\dagger {\mathcal H}_{\rm BdG}(\vec{k}) \Psi^{\rm Nam}_{\vec{k}}$, where $\Psi^{\rm Nam}_{\vec{k}}=\left[\Psi_{\vec{k}}, \tau_1 \Psi^\star_{-\vec{k}} \right]^\top$ is the Nambu-doubled spinor and 
\begin{equation}~\label{eq:Nambu_ham}
\mathcal{H}_{\rm BdG}(\vec{k}) =  d_1(\vec{k}) \Gamma_{01} + d_2(\vec{k}) \Gamma_{02} + d_3 (\vec{k}) \Gamma_{03} + \Delta \Gamma_{13}.
\end{equation}
The $4 \times 4$ Dirac matrices are $\Gamma_{ab}=\eta_a \otimes \tau_b$. The new set of Pauli matrices $\{ \eta_a \}$ act on the Nambu space. The factor of $1/2$ in $H_{\rm BdG}$ stems from the Nambu doubling.

Computation of the phase diagram of $\mathcal{H}_{\rm BdG}(\vec{k})$ is greatly simplified by noting that a unitary rotation by $U=\exp[-i \pi \Gamma_{20}/4]$ brings it to a block-diagonal form $U^\dagger \mathcal{H}_{\rm BdG}(\vec{k}) U=\mathcal{H}^+_{\rm BdG}(\vec{k}) \oplus \mathcal{H}^-_{\rm BdG}(\vec{k})$, where $\mathcal{H}^\pm_{\rm BdG}(\vec{k})={\boldsymbol \tau} \cdot \vec{d}^{\pm}(\vec{k})$ with $\vec{d}^\pm(\vec{k})=(d_1,d_2,d^\pm_3)(\vec{k})$ and
\begin{equation}
d^\pm_3 (\vec{k})=m_0 \pm \Delta -t_0[\cos(k_x a)+\cos(k_ya)]. 
\end{equation} 
The global phase diagram of $\mathcal{H}_{\rm BdG}(\vec{k})$ can now be constructed in terms of the total Chern number $C_{\rm tot}=C_+ + C_-$ as shown in Fig.~\ref{fig:pd_thc}(a), where $C_\pm$ are the Chern numbers for $\mathcal{H}^\pm_{\rm BdG}(\vec{k})$, computed from Eq.~\eqref{eq:chernnumber}. It features TSCs with $C_{\rm tot}=\pm 1$ and $\pm 2$, besides the ones with $C_{\rm tot}=0$. The $C_{\rm tot}=0$ sector fragments into two classes, which can be distinguished in terms of a weak topological invariant, namely, the Zak phase~\cite{Zak1989, Resta1994, Liu2017, Wu2020}. The one with a nontrivial Zak phase is named weak TSC~\cite{SM}.

\textbf{\textit{Thermal Hall effect}}.~We now compute responses of the paired states from Fig.~\ref{fig:pd_thc}(a), capturing the signatures of their nontrivial topological invariants. At this point, we should note that once superconductivity develops in the system, electrical charge responses become ill-defined as Cooper pairs do not obey the charge conservation. However, as the energy of the system is conserved, $\kappa_{xy}$ serves as a bona fide topological response to characterize the paired states. Details of the computation of $\kappa_{xy}$ in a six-terminal geometry has already been discussed. So, here we only quote the final results. We find that $\kappa_{xy}$ is nonvanishing only when $C_{\rm tot}$ is nonzero and half-integer quantized, namely, $\kappa_{xy}/\kappa_0=-C_{\rm tot}/2$~\cite{Read2000, Long2011, Fulga2020}. Therefore, TSCs with $C_{\rm tot}=\pm 1$ and $\pm 2$, give $\kappa_{xy}/\kappa_0=\mp 0.5$ and $\mp 1$, respectively, as shown in Fig.~\ref{fig:pd_thc}(b). However, $\kappa_{xy}=0$ whenever $C_{\rm tot}=0$, irrespective of whether the superconducting phase possesses a nontrivial Zak phase or not. It should be noted that the sign of $\kappa_{xy}$ can be changed without altering the nature of the TSC, namely the BdG band inversion momentum (${\bf K}^{\rm BdG}_{\rm inv}$), by taking ${\boldsymbol \tau} \to -{\boldsymbol \tau}$, for example. Thus, a full characterization of TSCs also demands a smoking gun probe of ${\bf K}^{\rm BdG}_{\rm inv}$.

In addition, we compute the longitudinal thermal conductance $G_{th}=(R^{\rm th}_{xx})^{-1}$, where $R^{\rm th}_{xx}=(T_3-T_2)/I_{\rm th}$ in the six-terminal setup. In TSCs with nontrivial $C_{\rm tot}$, $G_{th}=|k_{xy}|$, whereas $G_{th}=0$ in the trivial paired state. Most importantly, in the weak TSC phase $G_{th}/\kappa_0=1$. Therefore, $G_{th}$ always measures the number of edge modes equals to $2 (G_{th}/\kappa_0)$. Both (half)-quantized $G_{th}$ and $\kappa_{xy}$ are robust against random charge impurities of moderate strengths, except in the weak TSC phase where $G_{th}=\kappa_0$ survives only in the weak disorder regime~\cite{SM}.

\textbf{\textit{Edge band structure}}.~The topological nature of the superconductors and the associated ${\bf K}^{\rm BdG}_{\rm inv}$ can be established from the band structure of ${\mathcal H}_{\rm BdG}(\vec{k})$ in a semi-infinite system with only $k_x$ or $k_y$ as a good quantum number. One-dimensional $|C_{\rm tot}|$-fold degenerate edge modes then appear as dispersive states along $k_x$ or $k_y$, separated from the bulk states. See Fig.~\ref{fig:bs}. Furthermore, the edge modes cross the zero energy exactly at $\textbf{K}^{\rm BdG}_{\rm inv}$. We find that TSC with $C_{\rm tot}=-2$ ($+2$) supports doubly-degenerate edge states with the BdG band inversion at the $\Gamma$ (${\rm M}$) point. The $C_{\rm tot}=\pm 1$ TSCs replicate this outcome. But the edge modes are, then, non-degenerate. The paired state with $C_{\rm tot}=0$  supports counter-propagating edge modes, crossing the zero energy at $k_x$ or $k_y=0$ and $\pi$, only when it possesses a nontrivial Zak phase. Next we show that dislocation lattice defects probe ${\bf K}^{\rm BdG}_{\rm inv}$.

\textbf{\textit{Edge dislocation}}.~Two-dimensional edge dislocations are constructed from the so-called Volterra cut-glue procedure. The main idea is to cut a line of atoms up to a site, called the dislocation core as a first step. Subsequently, the sites across the cut are glued. This way, the system regains translational symmetry everywhere except near the dislocation core, where the missing translation characterizes the defect in terms of the Burgers vector ($\textbf b$). Due to this, when a BdG fermion encircles the defect core, it picks up a hopping phase $\exp[i \Phi_{\rm dis}]$, governed by the $\textbf{K} \cdot \textbf{b}$ rule~\cite{Ran2009, Teo2010, Asahi2012, Juricic2012, Hughes2014, Slager2012, Queiroz2019, Roy2021, Nag2021, Panigrahi2022NH, Panigrahi2022, Das2022}, where $\Phi_{\rm dis}= \textbf{K}^{\rm BdG}_{\rm inv} \cdot \textbf{b}$ (modulo $2\pi$). Following this principle, we find that TSCs with $C_{\rm tot}=-1$ ($-2$) support one (two) pair(s) of zero-energy dislocation modes. Furthermore, the TCS with $C_{\rm tot}=0$, but a nontrivial Zak phase features two zero-energy defect modes. See Fig.~\ref{fig:ldos}. In all these phases $\Phi_{\rm dis}=\pi$ (nontrivial) when $\textbf{b}= a \hat{{\bf e}}_{x}$ or $a \hat{{\bf e}}_{y}$, as $\textbf{K}^{\rm BdG}_{\rm inv}=(\pi,\pi)$ therein, resulting in edge modes crossing the zero energy at $k_x$ or $k_y=\pi$ [Fig.~\ref{fig:bs}]. For all the other paired states $\Phi_{\rm dis}=0$ (trivial). None of them, thus, hosts any zero-energy dislocation mode.

These observations can be supported from an alternative explanation. Note that two edges, introduced during the cut procedure, support counter-propagating edge modes. Once these two edges are glued, the associated edge modes hybridize and suffer level repulsion. When $n$ number of edge modes cross the zero energy at momentum $\pi$ or $0$, such a level repulsion can be modeled by a domain wall or uniform Dirac mass, acting on the edge subspace. Then, the Jackiw-Rebbi mechanism applies~\cite{Jackiw1976}, and in the former situation the dislocation core supports $n$ pairs of localized Majorana zero modes.

\textbf{\textit{Discussions}}.~From a paradigmatic toy square lattice model, featuring ${\mathcal T}$-breaking TIs with distinct topological invariant ($C$) and ${\bf K}^{\rm TI}_{\rm inv}$, here, we argue that their proximal NIs with band gap minima at ${\bf K}^{\rm NI}_{\rm min}={\bf K}^{\rm TI}_{\rm inv}$ can be distinguished, but only when TSCs develop in the system. In particular, $\kappa_{xy}$ and the response to the dislocation lattice defects inside the paired states (governed by ${\bf K}^{\rm BdG}_{\rm inv}={\bf K}^{\rm NI}_{\rm min}$) unambiguously distinguish parent NIs with different ${\bf K}^{\rm NI}_{\rm min}$'s. A generalization of this proposal possibly rests on the answer to the following question.

\emph{Can two NIs realized in the limits $m_0 \to \pm \infty$ be adiabatically connected?} In these two limits, the kinetic energy becomes unimportant and NIs can be modeled by a simple Hamiltonian ${\mathcal H}_{\rm NI}= m_0 \Gamma_{2N}$, where $\Gamma_{2N}$ is a $2N$-dimensional traceless Hermitian matrix and $2N$ is the total number of bands in the system, with $N=1$ in our model. The mass term ${\mathcal H}_{\rm NI}$ is always accompanied by a single Hermitian matrix as it does not break any fundamental or discrete lattice symmetry and, thus, transforms under the trivial singlet $A_{1g}$ representation under the crystallographic space group~\cite{tenfold2010, tenfoldPRB2008}. At half-filling, there are $N$ filled valence and $N$ empty conduction bands, with a band gap $2 m_0$ between them. Irrespective of the representation, eigenvalues of $\Gamma_{2N}$ are $+1$ and $-1$ (named generalized parity eigenvalues), and each of them is $N$-fold degenerate. The corresponding wave functions are parity eigenstates. When $m_0 \to \infty$, the conduction (valence) band is constituted by positive (negative) parity eigenstates. In the $m_0 \to - \infty$ limit, the situation is exactly the opposite. See, for example, Fig.~\ref{fig:pd}(a). As the parity eigenstates are orthogonal to each other, two atomic insulators realized in the limits $m_0 \to \pm \infty$, therefore, cannot be smoothly deformed into each other. This proof allows us to, at least, conjecture that our proposal to distinguish trivial atomic insulators by inducing TSCs should be applicable to systems of arbitrary dimensionality (above one) belonging to arbitrary symmetry class, as long as it can support distinct TIs with different ${\bf K}^{\rm TI}_{\rm inv}$'s. A further rigorous mathematical proof of this statement (if exists) is beyond the scope of the present Letter. See, however, Refs.~\cite{bernevig2017, Vishwanath2017NatComm}.

\textbf{\textit{Outlook}}.~Nature harbors a plethora of TIs with the hallmark band inversion at various points in the BZ~\cite{kanemele2006, BHZ2006, Fukane2007, moorebalents2007, rahulroy2009, tenfold2010, tenfoldPRB2008, vanderbilt2011, Fu2011, Slager2012, Shiozaki2014, BBH2017, BBHPRB2017, bernevig2017, Vishwanath2017NatComm, ashvin2018, Calugaru2019, Zhang2019, Vergniory2019, Tang2019, khalaf2021}. In these systems, TI-NI quantum phase transitions can be triggered by changing the quantum well width~\cite{quantumwell1} or via chemical substitutions~\cite{TINIQPT1, TINIQPT2, TINIQPT3, TINIQPT4, TINIQPT5} or by applying a hydrostatic pressure~\cite{hydropress1, hydropress2}. When doped, these quantum materials typically accommodate TSCs~\cite{TSCreview1}. Here, only for the sake of simplicity, we set the chemical potential to zero. Our proposal holds even when the insulators are doped, which favors nucleation of TSCs by forming a Fermi surface. Most importantly, doping lowers the threshold pairing amplitude to realize TSC and when the attractive pairing interaction resides only in the close proximity to the Fermi surface (the BCS pairing mechanism), realized within the valence or conduction band upon doping the insulators, TSCs appear for an \emph{infinitesimal} pairing amplitude~\cite{SM, RoyHOTSCSolo2020PRB}, making our proposal operative even away from the TI-NI critical point. Inclusion of longer-range hopping in the normal state often accommodates crystalline topological phases~\cite{Slager2012}, without removing the NIs with the band minima near the $\Gamma$ and ${\rm M}$ points nor the candidate TSC, promoting our proposal beyond the paradigm of toy models. Therefore, the task is to induce TSCs in doped topological materials with different ${\bf K}^{\rm NI}_{\rm inv}$'s after driving the system into a NI. While the thermal Hall conductivity is intimately tied with the breaking of the ${\mathcal T}$ in the paired state (class D), responses to dislocation lattice defects are applicable across all symmetry classes. Although challenging, $\kappa_{xy}$ nowadays is routinely measured with extremely high accuracy~\cite{ThermalHall1, ThermalHall2, ThermalHall3, ThermalHall4, ThermalHall5, ThermalHall6}, and Majorana dislocation modes can be detected via scanning tunneling microscope~\cite{dislocationSTM1, dislocationSTM2, Grainboundary2020NanoLetter}. Therefore, our proposal to distinguish NIs from the responses of their proximal TSCs can be tested in well-characterized topological quantum materials with existing experimental tools. Despite abundance of topological materials with the inversion symmetry in nature, which is only an emergent symmetry in our Letter, it will be worth an attempt to extend the jurisdiction of our proposal to systems where the inversion symmetry is broken at the microscopic level.

\textbf{\textit{Acknowledgments}}.~S.K.D was supported by a Startup Grant of B.R. from Lehigh University. S.M. was supported by Weizmann Institute of Science, Israel Deans fellowship through Feinberg Graduate School and the Raymond and Beverly Sackler Center for Computational Molecular and Material Science at Tel Aviv University. B.R.\ was supported by NSF CAREER Grant No.\ DMR- 2238679. We thank Suvayu Ali for technical support.

\bibliography{ref}

\end{document}